\documentclass{article}

\usepackage[final,nonatbib]{neurips_2020}

\usepackage[utf8]{inputenc} 
\usepackage[T1]{fontenc}    
\usepackage{hyperref}       
\usepackage{url}            
\usepackage{booktabs}       
\usepackage{amsfonts}       
\usepackage{nicefrac}       
\usepackage{microtype}      

\usepackage{amssymb,amsmath}
\usepackage{graphicx}
\usepackage{enumitem}
\usepackage{ulem}
\usepackage[small,bf]{caption}
\usepackage{xcolor}
\definecolor{gray}{rgb}{0.5, 0.5, 0.5}

\usepackage{siunitx}
\usepackage{subcaption}
\usepackage{multirow}
\usepackage{makecell}
\usepackage{wrapfig}
\usepackage{arydshln}

\title{Cross-Modality Protein Embedding for \\Compound-Protein Affinity and Contact Prediction}

\author{%
  Yuning You, Yang Shen\\
  Texas A\&M University\\
  \small{\texttt{\{yuning.you,yshen\}@tamu.edu}} 
  }

\begin{document}

\maketitle

\begin{abstract}

Compound-protein pairs dominate FDA-approved drug-target pairs and the prediction of compound-protein affinity and contact (CPAC) could help accelerate drug discovery. In this study we consider proteins as multi-modal data including 1D amino-acid sequences and (sequence-predicted) 2D residue-pair contact maps.  We empirically evaluate the embeddings of the two single modalities in their accuracy and generalizability of CPAC prediction (i.e. structure-free interpretable compound-protein affinity prediction).  And we rationalize their performances in both challenges of embedding individual modalities and learning generalizable embedding-label relationship.   We further propose two models involving  cross-modality protein embedding and establish that the one with cross interaction (thus capturing correlations among modalities) outperforms SOTAs and our single modality models in affinity, contact, and binding-site predictions for proteins never seen in the training set.  

\end{abstract}

\section{Introduction}

Computational prediction of compound--protein interactions (CPI) has been of great interest partly due to its potential impact on accelerating drug discovery \cite{kola2004can,paul2010improve}. Recent progress in this topic includes (1) the improved accuracy of structure-based binary classification \cite{lim2019predicting,Altman2019} and affinity regression \cite{gomes2017atomic,Jimenez2018} for CPI; (2) the structure-free inputs that remove the demand of compound-protein co-crystal or docked structures that are  experimentally or computationally expensive \cite{ozturk2018deepdta,gao2018interpretable,karimi2019deepaffinity,tsubaki2019compound,jiang2020drug}; and (3) the recent development of interpretable structure-free predictions of both protein-ligand binding affinities and their atomic contacts \cite{karimi2019deepaffinity,karimi2019explainable,li2020monn}.


We focus on interpretable CPI prediction without the need of compound-protein co-crystal or docked structures. Even unbound structures of proteins are not assumed here.  Specifically, we aim at simultaneous prediction of compound-protein affinity and contacts in the aforementioned structure-free setting.  We note that earlier works for this task represent proteins as 1D amino-acid sequences \cite{karimi2019explainable,li2020monn} or 1D structurally-annotated sequences \cite{karimi2019deepaffinity}. However, 1D sequences of proteins adopt 3D structures to function, including interactions with compounds; so structure-aware  representations of proteins (such as sequence-predicted residue-residue 2D contact maps) can also be useful, as explored in a recent affinity predictor \cite{jiang2020drug}.  (Although compound data can be available in both modalities of 1D SMILES and chemical graphs, we did not pursue both modalities and only represented compounds as graphs because SMILES strings have limited descriptive power and known worse performance in the CPAC task \cite{karimi2019deepaffinity,karimi2019explainable}.)


In this paper, we treat protein data as available in both modalities of 1D sequences and (sequence-predicted) 2D contact maps. And we ask the following questions: How do the two modalities compare with each other for the task of structure-free interpretable CPI prediction, i.e., compound-protein affinity and contact (CPAC) prediction? Is there an advantage to exploit  both modalities? And what would be a  beneficial cross-modality approach?  Our contributions and findings include the following: 
\begin{itemize}
    \item By embedding either modality with recurrent or graph neural networks and predicting affinities through intermolecular contact-predicting joint attentions, we empirically compared the two resulting single-modality models and found that: the 1D or 2D modality of proteins did not dominate each other for proteins seen in the training set; however, the 1D and 2D modality-based models tend to generalize better for unseen proteins in affinity prediction and contact prediction, respectively. We further provided conjectures involving the difficulty of embedding various modality and the mappings between various embeddings and affinity or contact labels.  
    \item For the first time, we propose cross-modality learning models for the task of structure-free interpretable CPI prediction, to capture and fuse the different information from both 1D \& 2D modalities of proteins.  And we empirically demonstrate that the two cross-modality learning models (through concatenation or cross-interaction of sequence and graph embeddings) achieve better accuracy and generalizability compared to the state of the art (SOTA) and our single-modality models, in compound-protein affinity, contact, and binding-site prediction.  
\end{itemize}


\section{Pipeline Overview}
\label{sec:pipeline}
\begin{wrapfigure}[15]{r}{0.5\linewidth}
\centering
\includegraphics[width=1\linewidth]{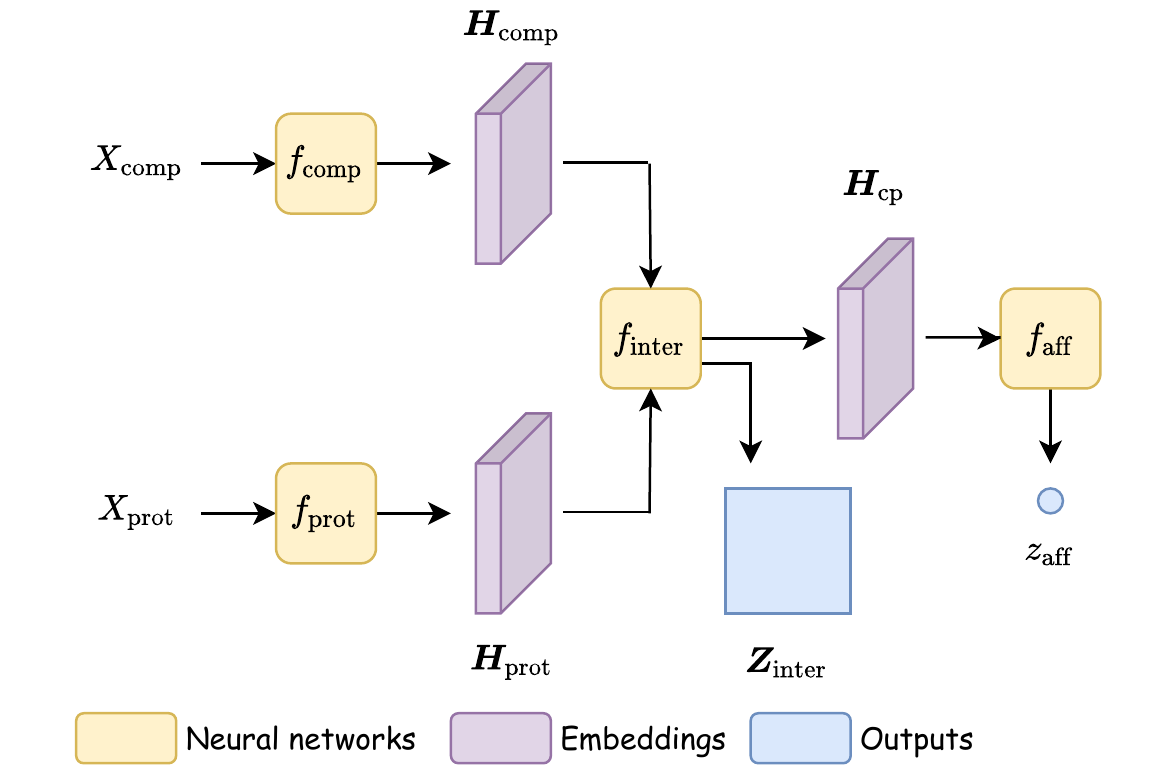}
\caption{Pipeline overview for compound-protein affinity and contact prediction model $f_\text{CPAC}$.}
\label{fig:pipeline}
\end{wrapfigure}

We assume that compounds are available in (1D SMILES or) 2D chemical graphs and proteins available in 1D amino-acid sequences.
Given a compound-protein pair $( X_{\text{comp}}, X_{\text{prot}} )$ composed of $N_{\text{comp}}$ atoms and $N_{\text{prot}}$ residues where $N_{\text{comp}}, N_{\text{prot}}$ are predefined and fixed numbers (padding is applied to ensure the fixed sizes), a CPAC model $f_\text{CPAC}: \mathbb{X}_{\text{comp}} \times \mathbb{X}_{\text{prot}} \rightarrow \mathbb{R}_{\ge 0} \times [0,1]^{N_{\text{comp}} \times N_{\text{prot}}}$ is targeted at making prediction for both the intermolecular affinity $z_\text{aff}$ and (atom-residue) contacts  $\boldsymbol{Z}_\text{inter}$,
where $\mathbb{X}_{\text{comp}}, \mathbb{X}_{\text{prot}}$ are respectively the spaces for $X_{\text{comp}}, X_{\text{prot}}$.
The SOTA pipelines for CPAC \cite{karimi2019explainable,karimi2019deepaffinity,li2020monn} comprise of the following three major components as shown in Figure \ref{fig:pipeline}.

(1) \textbf{Neural-network encoders} $f_{\text{comp}}: \mathbb{X}_{\text{comp}} \rightarrow \mathbb{R}^{N_{\text{comp}} \times D}, f_{\text{prot}}: \mathbb{X}_{\text{prot}} \rightarrow \mathbb{R}^{N_{\text{prot}} \times D}$ that separately extract embeddings  $\boldsymbol{H}_{\text{comp}}, \boldsymbol{H}_{\text{prot}}$ for the compound $X_{\text{comp}}$ and protein $X_{\text{prot}}$ where $D$ is hidden dimension.
Graph neural network (GNN, \cite{velivckovic2017graph,you2020l2,you2020does,you2020graph,liu2020towards,jin2020adversarial}) is adopted for compound 2D chemical graphs and hierarchical recurrent neural network (HRNN, \cite{el1996hierarchical}) is chosen for protein 1D amino-acid sequences.

(2) \textbf{Interaction module} $f_{\text{inter}}: \mathbb{R}^{N_{\text{comp}} \times D} \times \mathbb{R}^{N_{\text{prot}} \times D} \rightarrow [0, 1]^{N_{\text{comp}} \times N_{\text{prot}}} \times \mathbb{R}^{L \times D}$ taking the encoded embeddings $\boldsymbol{H}_{\text{comp}}, \boldsymbol{H}_{\text{prot}}$ as inputs,
employing joint attention to output the interaction matrix $\boldsymbol{Z}_\text{inter}$ and joint embedding to extract embeddings  $\boldsymbol{H}_{\text{cp}}$ for compound-protein pairs, where $L$ is hidden length determined by $N_{\text{comp}}, N_{\text{prot}}$.

(3) \textbf{Affinity module} $f_{\text{aff}}: \mathbb{R}^{L \times D} \rightarrow \mathbb{R}$ that predicts the affinity $z_{\text{aff}}$ given the joint embedding $\boldsymbol{H}_{\text{cp}}$, consisting of 1D convolutional, pooling layers, and multi-layer perceptron (MLP). Note that the contact-predicting interaction module feeds the affinity module, making affinity prediction intrinsically interpretable by the underlying contacts.  

After the CPAC model $f_\text{CPAC}$ forwardly generates the outputs $(z_{\text{aff}}, \boldsymbol{Z}_{\text{inter}})$, true labels $(y_{\text{aff}}, \boldsymbol{Y}_{\text{inter}})$ are compare to calculate the loss, $l_{\text{CPAC}}$, which consists of affinity loss $l_{\text{aff}}$, intermolecular atom--residue contact/interaction loss $l_{\text{inter}}$ and three structure-aware sparsity regularization loss $l_{\text{group}}, l_{\text{fused}}, l_{\text{L1}}$ described in \cite{karimi2019explainable}, expressed as:
\begin{equation}
    l_{\text{CPAC}} = l_{\text{aff}} + \lambda_{\text{inter}} l_{\text{inter}} + \lambda_{\text{group}} l_{\text{group}} + \lambda_{\text{fused}} l_{\text{fused}} + \lambda_{\text{L1}} l_{\text{L1}}. 
\end{equation} 
The model is trained end to end while the training loss is minimized.  More details can be found in
\cite{karimi2019explainable}.

\section{Single-Modality Models and Performances}
\label{sec:modaloty_comparison}

\textbf{Protein 1D sequences.}  We follow DeepAffinity+ \cite{karimi2019explainable} as described above and use HRNN to encode protein sequences. One change we made was 
replacing the hierarchical joint attention with na\"ive joint attention in the interaction module expressed as:
\begin{equation}
    \boldsymbol{Z}_\text{inter} = \boldsymbol{Z}'_\text{inter} / \mathrm{sum}(\boldsymbol{Z}'_\text{inter}), \quad z'_{\text{inter}, i,j} = (\boldsymbol{h}_{\text{comp},i} \boldsymbol{W}_{\text{comp},\text{attn}})^\mathrm{T} (\boldsymbol{h}_{\text{prot},j} \boldsymbol{W}_{\text{prot},\text{attn}}),
\end{equation}
where $z_{i,j} = \boldsymbol{Z}[i,j], \boldsymbol{h}_i = \boldsymbol{H}[i, :], i=1,...,N_\text{comp}, j=1,...,N_\text{prot}$ and $\boldsymbol{W}_{\text{comp},\text{attn}}, \boldsymbol{W}_{\text{prot},\text{attn}}$ are two learnable attention matrices.

\textbf{Protein 2D contact maps.}
In previous SOTAs for CPAC, proteins are often represented as 1D amino-acid sequences \cite{karimi2019explainable,karimi2019deepaffinity,li2020monn}.
We propose to adopt the 2D modality of proteins as inputs and model them as graphs with the following reasons.
Firstly, graphs are structure-aware compared with 1D sequences, potentially resulting in better generalizability. Secondly, graphs are more concise yet informative (focusing on pairwise residue interactions) compared to the data structure of 3D coordinates (which are also harder to predict than contact maps) \cite{Cao2020}.   
Lastly, the recent surge of development in graph  learning \cite{velivckovic2017graph,you2020l2,you2020does} provides advanced tools to facilitate graph representation learning.

Thus, given 2D residue-residue contact maps, we represent a protein input $X_{\text{prot}}$ as a graph $\mathcal{G}_{\text{prot}} = \{ \mathcal{V}_{\text{prot}}, \mathcal{E}_{\text{prot}} \}$ where vertices stand for residues and edges exist between residues predicted to be in contact (Z-scores of predicted probability are above 3).  When actual protein contact graphs are used for comparison, the edge criteria (for residue pairs in contact) is if their $C_\beta$ atoms are within 8\AA.  As the graphs are defined by the 2D contact maps, we may refer to them as 2D maps or 2D graphs interchangeably.



The graphs are associated with feature matrix $\boldsymbol{F}_{\text{prot}} \in \mathbb{R}^{N_{\text{prot}} \times D}$ (embedded amino-acid types of residues)
and the adjacency matrix $\boldsymbol{A}_{\text{prot}} \in \{0, 1\}^{N_{\text{prot}} \times N_{\text{prot}}}$ (binary contact map). 
We employ an expressive GNN model, graph attention network (GAT, \cite{velivckovic2017graph}) with $K$ layers as the protein encoder $f_{\text{prot}}$ to extract  graph embeddings, with the formulation of  each layer's forward propagation as:
\begin{align} \label{eq:gat}
    & \boldsymbol{H}_{\text{prot}}^{(k)} = \mathrm{MLP}(\tilde{\boldsymbol{S}}^{(k-1)} \boldsymbol{H}_{\text{prot}}^{(k-1)}), \quad
    \tilde{\boldsymbol{S}}^{(k-1)} = {\boldsymbol{D}^{(k-1)}}^{-1} (\boldsymbol{S}^{(k-1)} \odot \boldsymbol{A}_{\text{prot}}), \notag \\
    & \boldsymbol{S}^{(k-1)} = \mathrm{exp} (\boldsymbol{H}_{\text{prot}}^{(k-1)} \boldsymbol{W}^{(k-1)} {\boldsymbol{H}_{\text{prot}}^{(k-1)}}^\mathrm{T}),
\end{align}
where $\boldsymbol{H}_{\text{prot}} = \boldsymbol{H}_{\text{prot}}^{(K)}, \boldsymbol{H}_{\text{prot}}^{(0)} = \boldsymbol{F}_{\text{prot}}$, the normalization matrix $\boldsymbol{D}^{(k-1)} = \mathrm{diag}( (\boldsymbol{S}^{(k-1)} \odot \boldsymbol{A}_{\text{prot}}) \boldsymbol{J}_{N_{\text{prot}},1} )$, $\odot$ is the element-wise multiplication, $\boldsymbol{J}_{N_{\text{prot}},1}$ is an all-ones matrix with size $N_{\text{prot}} \times 1$, and $\boldsymbol{W}^{(k-1)}$ is a learnable weight matrix.

As (unbound or ligand-bound) structure  data is not readily available for many proteins, we use sequence-predicted 2D contact maps to overcome the limitation and broaden our models' applicability.  2D contact map prediction is done by RaptorX-contact \cite{wang2017accurate} that exploits both sequence and evolutionary information.

\textbf{Data set.} 
We use the dataset and splitting scheme as in DeepAffinity+ \cite{karimi2019explainable}, which is curated based on PDBbind \cite{liu2015pdb} and BindingDB \cite{liu2007bindingdb}.  It contains protein sequences, predicted (and actual bound) protein contact maps, compound SMILEs and graphs, affinity labels (p$K_d$/p$K_i$) and intermolecular atomic interactions/contacts (curated from the LigPlot service of PDBsum \cite{laskowski2018pdbsum}). The updated dataset is diverse: it consists of 4,446 pairs
between 3,672 compounds (of wide range of properties such as logP, molecular weight, and affinity labels) and 1,287 proteins (including enzymes across all six classes, GPCRs, nuclear receptors, ion channels, and so on). The dataset is split into subsets of various challenging levels in generalizability: 795 pairs involving unseen proteins (proteins not present in the training set), 521 pairs involving unseen compounds, and 205 for unseen both; whereas the rest is randomly split into training including validation (2,334) and the default test (591) sets.  Note that the default test set contains compounds or protein seen in the training set but never training compound-protein pairs.  


\textbf{Model training and hyperparameter tuning.}
We train our models end to end with the following optimization settings as in \cite{karimi2019explainable}: the optimizer Adam with a learning rate of 0.001, the batch size of 64 and the maximum  amount of training epochs being 200.  The best checkpoint model is selected via validation.
The following hyperparameters in the loss function are optimized following a two-stage process over pre-defined grids \cite{karimi2019explainable}.  Specifically,  $\lambda_\text{group}$, $\lambda_\text{fused}$, and $\lambda_\text{L1}$ are first tuned over $\{ 0.01, 0.001, 0.0001 \}$ with $\lambda_\text{inter} = 0$ (affinity regression alone), where the best affinity loss $l_\text{aff}$ is recorded and $\lambda_\text{group}$, $\lambda_\text{fused}$, and $\lambda_\text{L1}$ are optimized  with the best AUPRC such that the corresponding affinity RMSE does not deteriorate more than  10\% of the best affinity RMSE. In the second stage, we fix the optimal $\lambda_\text{group}$, $\lambda_\text{fused}$, and  $\lambda_\text{L1}$ and tune  $\lambda_\text{inter}$ over $\{ 1\mathrm{e}0, 1\mathrm{e}1, 1\mathrm{e}2, 1\mathrm{e}3, 1\mathrm{e}4, 1\mathrm{e}5 \}$  based on the best AUPRC performance while jointly optimizing the regularized affinity and contact losses.

\textbf{Numerical comparison of different modalities.}
We compare the empirical results in Table \ref{tab:deepaffinity_plusplus} between taking 1D amino-acid sequences  and 2D contact maps as protein inputs, using HRNN and GAT as encoders for proteins, respectively.  
We make the following observations.
\begin{table}[!htb]
\scriptsize
\begin{center}
\caption{Affinity and contact prediction with different modalities of proteins as inputs.}
\label{tab:deepaffinity_plusplus}
\begin{tabular}{c | c | c c | c c}
    \hline
    \hline
    \multicolumn{2}{c|}{ } & \multicolumn{2}{c|}{1D Sequences} & \multicolumn{2}{c}{2D Graphs} \\ \cline{3-6}
    \multicolumn{2}{c|}{ } & Test (Seen-Protein) & Unseen-Protein & Test (Seen-Protein) & Unseen-Protein \\
    \hline
    \hline
    Affinity & RMSE $\downarrow$ & 1.57 & 1.63 & 1.49 & 1.75 \\
    Prediction & Pearson's $r$ $\uparrow$ & 0.67 & 0.44 & 0.68 & 0.43 \\
    \hdashline
    Contact & AUPRC (\%) $\uparrow$ & 20.51 & 6.54 & 17.29 & 8.78 \\
    Prediction & AUROC (\%) $\uparrow$ & 79.01 & 73.03 & 77.34 & 77.94 \\
    \hline
    \hline
\end{tabular}
\end{center}
\end{table}

(i) For affinity prediction (see RMSE \& Pearson), 
1D sequences and 2D graphs did not yield major differences especially in Pearson's $r$.  1D sequences led to less  deterioration in RMSE from the validation set (containing seen proteins) to unseen proteins.  

One conjecture is that the information in graphs might be more difficult to learn compared to sequences (the training RMSE losses are 0.71 \& 0.99 for 1D \& 2D modalities, respectively, when long enough training processes were performed). 
Moreover, affinity prediction for unseen-protein cases are not as challenging as intermolecular contact prediction to show the benefit of the 2D modality (see (ii) below), as contact prediction often involves tens of thousands of values (rather than a single value) to fit for each compound-protein pair.   

(ii) For contact prediction (see AUPRC \& AUROC), encoding proteins as 1D sequences  performed better (+3.22\% at AUPRC and +1.67\% at AUROC) in seen proteins, (i.e. the proteins in compound-protein pairs at the inference phase are involved in the training compound-protein pairs).  
Meanwhile, encoding 2D protein contact maps (graphs) outperformed doing that to 1D protein sequences (+4.91\% at AUPRC and +2.24\% at AUROC) for unseen proteins. 

We conjecture that sequential dependency information encoded in 1D amino-acid sequences is well captured especially for seen proteins whose embeddings are well constructed after training (as they are already represented in the training set), leading to the better contact predictions for seen proteins. However, the sequential information learned from the encoder could be more accurate toward intermolecular contact prediction for close or even distant homologs of seen proteins but it is less general to unseen proteins.  

In contrast, we conjecture that the better generalizability of the 2D modality model might result from the quality of the encoded embedding of proteins, which is co-determined by both the inputs (2D maps) and encoders (GAT models). 
The structural topology information encoded in protein 2D contact maps is more difficult for graph neural networks to capture even for seen proteins, leading to the worse contact predictions for seen proteins.  However, such information can generalize to unseen proteins well toward contact prediction.  In particular, even when sequence similarity for non-homologous proteins (to training ones) is too low to be detectable using RNNs, binding-pocket (subgraph) similarity could still preserve and be detected in 2D contact maps using GNNs thus eventually  leads to better intermolecular contact prediction.  
\section{Cross-Modality Models}
We have shown that both sequential dependency in 1D amino-acide sequences and structural  topology in 2D contact maps are important information for proteins to extract accurate and generalizable embeddings.  
Therefore it is natural to propose a cross-modality learning framework 
 that captures and fuses the information from 1D \& 2D modalities for better performances. Specifically we have designed the following two models.

\begin{figure}[!htb]
\centering
\includegraphics[width=1.05\linewidth]{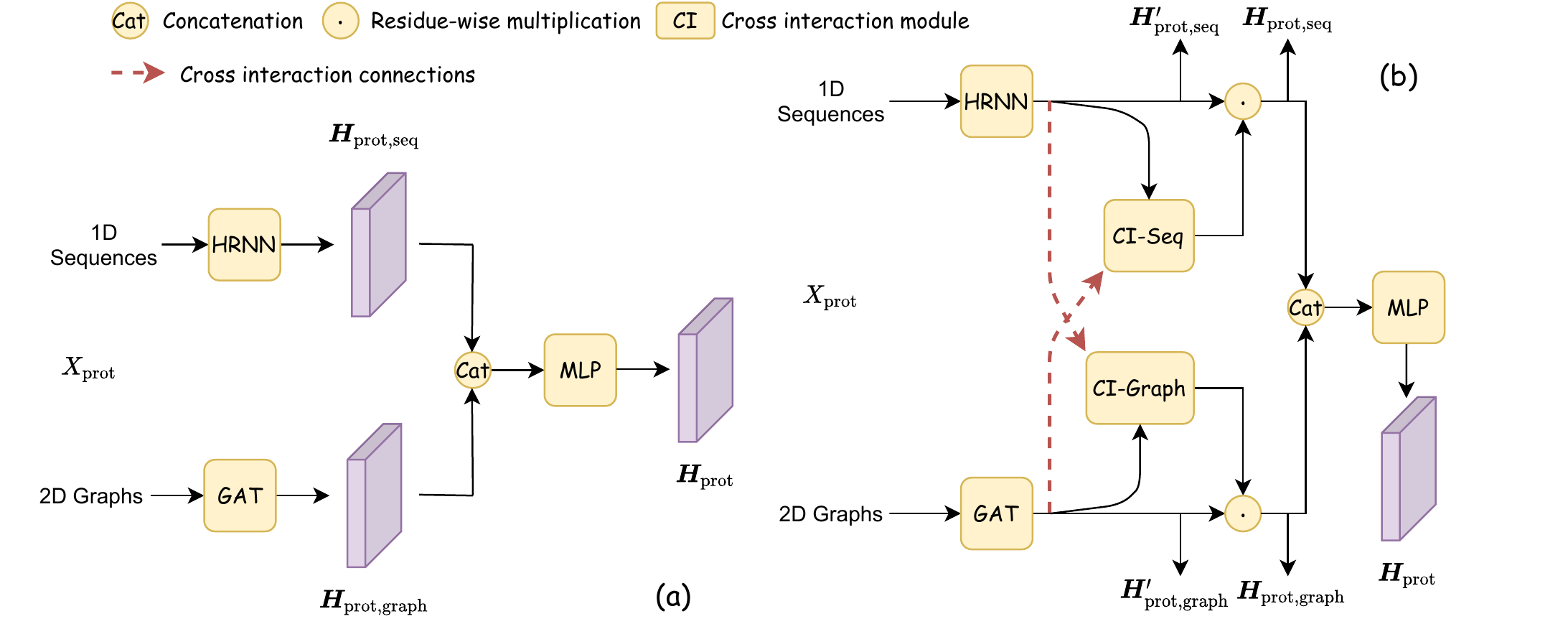}
\caption{Cross-modality encoder for proteins to capture and fuse different modality information, with (a) na\"ive concatenation and (b) cross interaction introduced.}
\label{fig:deepmodality}
\end{figure}

\textbf{Concatenation.}
A simple fusion model is to concatenate the extracted embeddings of the 1D and 2D modalities that are encoded by HRNN and GAT, respectively, as shown in Figure \ref{fig:deepmodality}(a).   
Indeed, concatenation is commonly used in previous work \cite{hamilton2017inductive,xu2018powerful} to preserve information from different sources.  The concatenated output is fed to a multi-layer perception (MLP) for the final protein embedding $\boldsymbol{H}_{\text{prot}}$.  

\textbf{Cross interaction.}
Although the aforementioned concatenation strategy preserves the information of individual modalities, the encoding processes for the two modalities are separate.  In other words, the two types of embeddings from different modalities were independently encoded and then mixed through concatenation.  However, the different modalities of proteins are intrinsically correlated with each other and could be coupled in a properly-designed representation-learning process.  Therefore, we have introduced a cross interaction module to facilitate the encoder to learn protein embeddings from correlated data (1D and 2D modalities), as shown in Figure \ref{fig:deepmodality}(b).
Specifically, given the outputs of encoders $\boldsymbol{H}'_{\text{prot,seq}}$ and $\boldsymbol{H}'_{\text{prot,graph}}$, we calculate sequence \& graph cross-modality outputs $\boldsymbol{H}_{\text{prot,seq}}$ and  $\boldsymbol{H}_{\text{prot,graph}}$, respectively:
\begin{align} \label{eq:cross_interaction}
    & \boldsymbol{h}_{\text{prot,seq},n} = (\mathrm{sigmoid} ({\boldsymbol{h}''_{\text{prot,graph},n}}^\mathrm{T} \boldsymbol{h}'_{\text{prot,seq},n}) + 1) \boldsymbol{h}'_{\text{prot,seq},n}, 
    \\
    & \boldsymbol{h}_{\text{prot,graph},n} = (\mathrm{sigmoid} ({\boldsymbol{h}''_{\text{prot,seq},n}}^\mathrm{T} \boldsymbol{h}'_{\text{prot,graph},n}) + 1) \boldsymbol{h}'_{\text{prot,seq},n},
\end{align}
where $\boldsymbol{h}_n^{\cdot} = \boldsymbol{H}^{\cdot}[n, :]$ ($\cdot$ can be empty, $'$ or $''$),  $\boldsymbol{H}''_{\text{prot,graph}} = \boldsymbol{H}'_{\text{prot,graph}} \boldsymbol{W}_{\text{cross,graph}}$, $\boldsymbol{H}''_{\text{prot,seq}} = \boldsymbol{H}'_{\text{prot,seq}} \boldsymbol{W}_{\text{cross,seq}}$, and $\boldsymbol{W}_{\text{cross,seq}}$ and $\boldsymbol{W}_{\text{cross,graph}}$ are learnable weight matrices.  
Instead of independently extracting information from protein modalities (1D sequences and 2D contact maps), the cross interaction module enforces a learned relationship between the encoded embeddings of the two protein modalities, which is expected to better capture the information from the correlated protein modalities and to benefit the affinity and contact prediction.  Again, $\boldsymbol{H}_{\text{prot,seq}}$ and  $\boldsymbol{H}_{\text{prot,graph}}$ (now with information from each other) are concatenated and fed to an MLP for the final protein embedding $\boldsymbol{H}_{\text{prot}}$. 

The idea of cross interaction was previously introduced in \cite{tan2019lxmert} and modified in our study as follows.  First, we do not normalize cross interaction along residues (sequence length is 1000 here) since it would significantly change the scale of the residue embeddings. Second, we restrict the cross interaction for each residue in the range of [0, 1] with $\text{sigmoid}$ function to represent the cross-modality ``interaction strength''. 



\section{Results}
We compare our single-modality and cross-modality models with two latest SOTAs for the CPAC problem, namely Gao et al. \cite{gao2018interpretable} and DeepAffinity+ \cite{karimi2019explainable}.  Tasks involved include affinity, contact, and binding-site predictions.

\textbf{Affinity and Contact Prediction.}
As shown in Table \ref{tab:sota} and \ref{tab:sota_contacts}, compared to SOTAs, our models have achieved similar performances in affinity prediction (RMSE and Pearson's $r$) and improved performances in contact prediction (AUPRC and AUROC) especially for proteins never seen in training (unseen-protein and unseen-both). We have made the following observations.

\begin{table*}[!htb]
\begin{center}
\caption{Comparison among SOTAs and our models in compound-protein affinity prediction (measured by RMSE and Pearson's correlation coefficient). $^*$ denotes the cited performances. Boldfaced were the best performances for given test sets.}
\label{tab:sota}
\resizebox{0.9\textwidth}{!}{
\begin{tabular}{c | c | c c c c}
\hline
\hline
& & Test (Seen-Both) & Unseen-Compound & Unseen-Protein & Unseen-Both \\
\hline
\hline
 \multicolumn{6}{c}{SOTAs} \\ \hline 
\multirow{2}{*}{Gao et al.$^*$} & RMSE & 1.87 & 1.75 & 1.72 & 1.79 \\
 & Pearson's $r$ & 0.58 & 0.51 & 0.42 & 0.42 \\
\hline
\multirow{2}{*}{DeepAffinity+$^*$} & RMSE & 1.49 & \textbf{1.34} & 1.57 & 1.61 \\
 & Pearson's $r$ & \textbf{0.70} & 0.71 & 0.47 & 0.52 \\
  \hline
 \multicolumn{6}{c}{Ours} \\
\hline
\multirow{2}{*}{\thead{Single Modality\\ (1D Sequences)}} & RMSE & 1.57 & 1.38 & 1.63 & 1.79 \\
 & Pearson's $r$ & 0.67 & \textbf{0.73} & 0.44 & 0.402 \\
\hline
\multirow{2}{*}{\thead{Single Modality\\ (Pred. 2D Graphs)}} & RMSE & 1.49 & 1.37 & 1.75 & 1.93 \\
 & Pearson's $r$ & 0.68 & 0.70 & 0.43 & 0.34 \\
\hline
\multirow{2}{*}{\thead{Single Modality\\ (True 2D Graphs)}} & RMSE & 1.69 & 1.62 & 1.88 & 1.99 \\
 & Pearson's $r$ & 0.59 & 0.58 & 0.33 & 0.25 \\
\Xhline{2.5\arrayrulewidth}
\multirow{2}{*}{\thead{Cross Modality\\ (Concatenation)}} & RMSE & \textbf{1.47} & 1.37 & 1.78 & 1.91 \\
 & Pearson's $r$ & 0.68 & 0.71 & 0.47 & 0.40 \\
\hline
\multirow{2}{*}{\thead{Cross Modality\\ (Cross Interaction)}} & RMSE & 1.55 & 1.43 & \textbf{1.56} & \textbf{1.62} \\
 & Pearson's $r$ & 0.65 & 0.68 & \textbf{0.50} & \textbf{0.53}  \\
\hline
\hline
\end{tabular}}
\end{center}
\end{table*}

\begin{table*}[!htb]
\begin{center}
\caption{Comparison among SOTAs and our models in contact prediction (measured by AUPRC and AUROC). $^*$ denotes the cited performances. Boldfaced were the best performances for given test sets.}
\label{tab:sota_contacts}
\resizebox{0.9\textwidth}{!}{
\begin{tabular}{c | c | c c c c}
\hline
\hline
& & Test (Seen-Both) & Unseen-Compound & Unseen-Protein & Unseen-Both \\
\hline
\hline
 \multicolumn{6}{c}{SOTAs} \\ \hline 
\multirow{2}{*}{Gao et al.$^*$} & AUPRC (\%) & 0.60 & 0.57 & 0.48 & 0.48 \\
 & AUROC (\%) & 51.57 & 51.50 & 51.65 & 51.55 \\
\hline
\multirow{2}{*}{DeepAffinity+$^*$} & AUPRC (\%) & 19.74 & 19.98 & 4.77 & 4.11 \\
 & AUROC (\%) &  73.78 & 73.80 & 60.01 & 59.09 \\
  \hline
 \multicolumn{6}{c}{Ours} \\
\hline
\multirow{2}{*}{\thead{Single Modality\\ (1D Sequences)}} & AUPRC (\%) & 20.51 & 20.80 & 6.54 & 6.36 \\
 & AUROC (\%) & 79.01 & 80.00 & 73.03 & 73.41 \\
\hline
\multirow{2}{*}{\thead{Single Modality\\ (Pred. 2D Graphs)}} & AUPRC (\%) & 17.29 & 17.46 & 8.78 & 7.05 \\
 & AUROC (\%) & 77.34 & 78.70 & 77.94 & 76.59 \\
\hline
\multirow{2}{*}{\thead{Single Modality\\ (True 2D Graphs)}} & AUPRC (\%) & 21.41 & 21.33 & 10.52 & 9.40 \\
 & AUROC (\%) & \textbf{84.60} & \textbf{85.17} & \textbf{84.08} & \textbf{84.29} \\
\Xhline{2.5\arrayrulewidth}
\multirow{2}{*}{\thead{Cross Modality\\ (Concatenation)}} & AUPRC (\%) & \textbf{23.85} & \textbf{23.52} & 7.74 & 7.29 \\
 & AUROC (\%) & 80.90 & 81.64 & 80.59 & 78.95 \\
\hline
\multirow{2}{*}{\thead{Cross Modality\\ (Cross Interaction)}} & AUPRC (\%) & 23.49 & 23.29 & \textbf{12.43} & \textbf{9.60} \\
 & AUROC (\%) & 81.30 & 82.07 & 80.64 & 79.78 \\
\hline
\hline
\end{tabular}}
\end{center}
\end{table*}

First, our models used similar backbone as  DeepAffinity+ and revised the joint attention mechanism; thus DeepAffinity+ and our 1D sequence-based single-modality model, both using protein sequences, had similar performances in affinity prediction but ours improved contact prediction. 

Second, as observed in Section \ref{sec:modaloty_comparison}, compared to the 1D modality of protein sequences, the 2D modality of (sequence-predicted) protein contact maps improved the generalizability of compound-protein contact prediction for unseen proteins or unseen both, even though it resulted in slightly worse accuracy for seen proteins.  Higher-quality actual protein contact maps, compared to sequence-predicted ones, further benefited  contact prediction for both seen and unseen proteins; but they could lead to worse affinity prediction.  These results echo our earlier conjecture that structural topology in the 2D graphs is more informative for the more complex task of contact prediction even though it may not be as effective as the 1D sequences for the less complex task of affinity prediction.  

We have also made the following observations for our cross-modality fusion models where only sequence-predicted protein contact maps are used.  

Third, fusing two modalities' information together, even by a simple concatenation strategy, could get the best of both modalities: the cross modality model by concatenation had better contact prediction than single-modality models (even the true 2D map-based one) and a trade-off in affinity predictions (better than the 2D single modality models and worse than the 1D single modality model). These results confirm our rationale of proposing cross-modality protein encoders for the CPAC task.  

Last, enforcing a learned correlation between the 1D and 2D embeddings rather than independently learning two individual embeddings, the cross-modality model with cross interaction further improved  affinity prediction and actually had the best affinity accuracy among all methods for unseen proteins or unseen both.  Moreover, it impressively achieved the best AUPRC for unseen proteins and unseen both.  We note that, as intermolecular contacts only represent a minority (around 0.4\%) of all compound-protein atom-residue pairs, AUPRC is a much more relevant measure than AUROC for contact prediction. These results reinforced our rationale that the learned correlation between embeddings from different modalities can better capture the correlated data and better perform CPAC predictions.

\textbf{Protein binding-site prediction.}  We also compare Gao et al., DeepAffinity+, and our models for protein binding site prediction that is ligand-specific and structure-free.  Our models again significantly improve the accuracy here compared to SOTAs.  As actual protein structures (unbound or bound) are not assumed available, the single-modality model using true 2D contact maps (from compound-bound protein structures) here is essentially providing an estimate of the performance upper bound for unseen proteins. Impressively, using only protein sequences and sequence-predicted contact maps, both cross-modality models improved against the single modality model (true 2D graphs) for seen proteins and performed closely to the latter for unseen proteins.  The cross-modality model with cross interaction achieved the best AUPRC for unseen proteins among all models compared.  Again, as  protein binding-site residues represent a minority among all residues, AUPRC is a much more relevant measure than AUROC for assessing binding-site prediction.  

\begin{table*}[!htb]
\begin{center}
\caption{Comparison among SOTAs and our models in ligand-specific and structure-free protein binding-site prediction. $^*$ denotes the cited numbers. Boldfaced are the best performances for individual test sets.}
\label{tab:binding_site}
\resizebox{0.9\textwidth}{!}{
\begin{tabular}{c | c | c c c c}
\hline
\hline
& & Test (Seen-Both) & Unseen-Compound & Unseen-Protein & Unseen-Both \\
\hline
\hline
\multicolumn{6}{c}{SOTAs} \\
 \hline 
\multirow{2}{*}{Gao et al.$^*$} & AUPRC (\%) & 5.43 & 5.38 & 4.95 & 4.96 \\
 & AUROC (\%) & 49.79 & 50.51 & 48.21 & 48.74 \\
\hline
\multirow{2}{*}{DeepAffinity+$^*$} & AUPRC (\%) & 42.16 & 43.14 & 16.98 & 15.65 \\
 & AUROC (\%) & 76.33 & 78.22 & 64.93 & 65.18 \\ \hline
 \multicolumn{6}{c}{Ours} \\
\hline
\multirow{2}{*}{\thead{Single Modality\\ (1D Sequences)}} & AUPRC (\%) & 40.35 & 40.81 & 20.37 & 20.17 \\
 & AUROC (\%) & 76.69 & 77.79 & 70.28 & 70.96 \\
\hline
\multirow{2}{*}{\thead{Single Modality\\ (Pred. 2D Graphs)}} & AUPRC (\%) & 33.17 & 33.83 & 25.57 & 22.49 \\
 & AUROC (\%) & 75.11 & 76.53 & 76.15 & 74.87 \\
\hline
\multirow{2}{*}{\thead{Single Modality\\ (True 2D Graphs)}} & AUPRC (\%) & 41.73 & 42.58 & 29.44 & \textbf{29.02} \\
 & AUROC (\%) & \textbf{83.67} & \textbf{84.85} & \textbf{83.82} & \textbf{84.15} \\
\Xhline{2.5\arrayrulewidth}
\multirow{2}{*}{\thead{Cross Modality\\ (Concatenation)}} & AUPRC (\%) & \textbf{43.56} & \textbf{44.12} & 28.15 & 26.44 \\
 & AUROC (\%) & 78.83 & 79.75 & 78.51 & 77.61 \\
\hline
\multirow{2}{*}{\thead{Cross Modality\\ (Cross Interaction)}} & AUPRC (\%) & 43.45 & 43.00 & \textbf{30.54} & 27.18 \\
 & AUROC (\%) & 78.85 & 79.73 & 77.37 & 77.54 \\
\hline
\hline
\end{tabular}}
\end{center}
\end{table*}

\section{Conclusions}

We explore in this study various protein modalities (1D sequences and 2D residue-residue contact maps) in the context of compound-protein affinity and contact prediction.  To this end, we have exploited RNNs and GNNs to encode the 1D and 2D modalities respectively and proposed cross-modality models (concatenation and cross interaction) on top of the single-modality models.  

Our experiments show that the two different protein modalities result in different accuracy and generalizability in affinity and contact predictions.  Specifically, sequential dependency learned in the 1D protein modality can be adequate for the relatively simple task of affinity prediction.  However, it does not generalize well for the relatively difficult task of contact prediction especially when the proteins are new. In other words, the accuracy of learned sequence-contact mapping can be restricted to seen proteins or their homologs but does not transfer to a non-homolog.  In contrast, structural topology in the 2D protein modality is more difficult to capture by GNNs and its mapping to affinity can be predicted less well (not to mention that the quality of the predicted 2D modality is worse than the actual).  However, once the mapping between the 2D embeddings and intermolecular contacts is learned, it generalizes well to unseen proteins, possibly due to better capturing  subgraph (binding pocket) similarity.   

Our experiments also show that cross-modality models can exploit the correlation between both modalities and enjoy the benefits of both modalities even when a simple concatenation strategy is adopted for the two embeddings.  The newly proposed cross interaction model has led to better affinity prediction (RMSE and Pearson's $r$) \text{and} better contact prediction (AUPRC) for unseen proteins than SOTAs, any our  single-modality model, and the simple cross-modality model with concatenation.  It has also outperformed those other models in the generalizability of binding-site prediction for unseen proteins.  

\section*{Acknowledgment}
This project is in part supported by the
National Science Foundation (CCF-1943008 to YS) and the National Institute of General Medical Sciences of the National Institutes of Health (R35GM124952 to YS).  
We thank Texas A\&M High Performance Research Computing (HPRC) for computing allocations.  We also thank anonymous reviewers for useful comments that have helped improve the manuscript.  

\bibliographystyle{unsrt}
\bibliography{ref}

\end{document}